\documentclass[11pt]{article}
\usepackage[T2A]{fontenc}
\usepackage[cp866]{inputenc}
\usepackage[english]{babel}
\usepackage[dvips]{graphicx }
\usepackage{amsmath,euscript,amssymb}
\newcommand{\be}{\begin{equation}}
\newcommand{\ee}{\end{equation}}
\newcommand{\bea}{\begin{eqnarray}}
\newcommand{\eea}{\end{eqnarray}}
\def\vh{\varphi}

\title{Bogoliubov's Integrals of Motion\\
in Quantum Cosmology and Gravity}

\author{V.N. Pervushin, and V.A. Zinchuk,\\
Bogoliubov Laboratory of Theoretical Physics,\\
Joint Institute for Nuclear Research, 141980 Dubna, Russia}

\begin{document}

\maketitle

\begin{abstract}

 Quantum Cosmology and Gravity are formulated here as the primary and
secondary quantizations of the energy constraints by analogy with
the historical formulation of  quantum field theory. New fact is
that  both the Universe and its matter are created from stable
vacuum obtained by  the Bogoliubov-type transformation just as it
is in the theory of  quantum superfluid liquid. Such the Quantum
Gravity gives us possibility  to explain topical problems of
cosmology by the cosmological creation of universes and particles
from vacuum.


\end{abstract}

\vspace{1cm}

 {\small  Proceedings of the II International Conference on
Super\-integrable Systems in Classical and Quantum Mechanics,
Dubna, Russia, June~27 - July 1, 2005, \rm will be published in
\it  Yadernaya Fizika \rm (2006). }

\vspace{1cm}

\tableofcontents

\newpage

\section{Introduction}

 Quantization of General Relativity (GR) \cite{einsh,H}
 and its cosmological approximation  is one of topical problems of modern physics.
 Wheeler and DeWitt \cite{WDW} proposed to formulate
  the quantum cosmology and gravity   by
 analogy  with the primary  quantization of the energy constraint
 in Special Relativity (SR)  \cite{Logunov} introducing
  the field space of events
 in GR, where the role of time-like variable is played by the spatial
 metric determinant  converting into
  the cosmological scale factor in the homogeneous approximation.

  In this case the main problem is  resolving  the Wheeler -- DeWitt
  (WDW)  equation. In the present paper, in order to solve this
  WDW equation and find its integrals of motion, we propose to use
  the holomorphic variables and their 
  Bogoliubov transformations \cite{B,ps1}.

 In Section 2,  the statement of problem
 and the direct way from Einstein -- Hilbert's geometric formulation
 of  classical cosmology to the quantum one are considered.
  Section 3 is devoted to a similar quantization of
 Einstein -- Hilbert's  GR in the spirit of
  the microscopic theory of superfluidity \cite{B,Lan,Lon}.

\section{Quantum Cosmology}

\subsection{Foundations of General Relativity}

  Einstein's GR \cite{einsh} is based on the dynamic action
   proposed
  by Hilbert \cite{H}
  \be\label{h1}
 S_{GR}=\int d^4x\sqrt{-g}
 \left[-\frac{\vh_0^2}{6}~~R(g)+{\cal L}_{\rm
 matter}\right],~~~~~\vh_0^2=\frac{3}{8\pi}M^2_{Planck},
 \ee
 where $M^2_{Planck}$ is the inverse coupling Newton constant in units
 $\hbar=c=1$.
  This action is given in a Riemannian
 space-time manifold with  {\it geometric interval}
 \be\label{h2}
 ds=g_{\mu\nu}dx^{\mu}dx^{\nu}.
 \ee
  Both the action and interval are invariant with respect
  to general coordinate transformations
  \be\label{gct}
 x^{\mu}~~~~~\longrightarrow~~~~~\tilde{x}^{\mu}=
 \tilde{x}^{\mu}(x^0,x^1,x^2,x^3).
 \ee

\subsection{Cosmology on the pathway to QFT}

 In order to demonstrate to a reader the direct pathway from
  Einstein -- Hilbert's ``Foundation of Physics'' \cite{einsh,H} to QFT of
  universes,
  we consider GR
 in the homogeneous approximation of the
 interval
 $$
 ds^2\simeq ds^2_{\rm WDW}=a^2(x^0)[(N_0(x^0)dx^0)^2 -(dx^idx^i)].
 $$
 where Hilbert's action  (\ref{h1}) and interval  (\ref{h2}) take the form
\be\label{H1}
 S_{\rm cosmic}=V_0\int
 dx^0N_0\left[-\left(\frac{d\varphi}{N_0dx^0}
\right)^2-\rho_0(\varphi)\right],  
 \ee
\be\label{conf}
 d\eta=N_0(x^0)dx^0~~~\longrightarrow~~
 ~~\eta=\int\limits_{0}^{x^0} d\overline{x}^0 N_0(\overline{x}^0),
 \ee
 here $a(x^0)$ is the  cosmological scale
 factor,
 \be\label{fi}\vh(x^0)=\vh_0a(x^0)\ee is the running scale of all masses,
 $\rho_0(\vh)$ is the energy density of the matter in a universe,
 $V_0$ is a spatial volume,
  and $N_0$ is the lapse function.
 This homogeneous approximation keeps the symmetry of the action
 and the  interval with respect to
  reparametrizations of the  coordinate evolution
 parameter $x^0~\to~\overline{x}^0=\overline{x}^0(x^0)$.
 Recall that similar diffeomorphisms  in the case of SR
 leads to energy constraint
  determining
 observable quantities of the type of energy, time-like variable and number of
 particles in quantum theory \cite{Logunov,pp}.

 In the WDW cosmology this gauge symmetry
 means that the scale factor $\vh$
 is the {\it time-like variable} in the {\it field space of events}
 introduced
 by Wheeler and DeWitt
 in \cite{WDW}, and the canonical momentum $P_\vh$ is the corresponding
 Hamiltonian that becomes the energy $E_\vh$ in equations of motion.

 The direct historical pathway from  SR to QFT of particles
 \cite{Logunov}
  shows us a similar direct pathway
  from  Hilbert's ``Foundation of Cosmology'' (\ref{H1}), (\ref{conf})
 to ``QFT'' of universes.
 Wheeler and DeWitt \cite{WDW} included in this pathway

1) the Hamiltonian approach: \be\label{ham}
 S_{\rm cosmic}=\int
 dx^0\left[-P_\vh\partial_0
 \vh-\frac{N_0}{4V_0}\left(-P^2_\vh+E^2_\varphi\right)\right],
 \ee
 where
 \be\label{ev1}
 E_\varphi=2V_0\sqrt{\rho_0(\vh)}
 \ee
 is treated as the {\it''energy of a universe''},

2) constraining: $P^2_{\varphi}-E^2_\varphi=0$,

3) primary quantization:
$[\hat{P}^2_{\varphi}-E^2_\varphi]\Psi=0$, here
$\hat{P}_{\varphi}=-id/d\varphi$.

\noindent We proposed here to include also the transition to
holomorphic variables \cite{ps1} used in

 4) secondary quantization: $\Psi=([A^++A^-]/\sqrt{2E_\vh})$,

5)  Bogoliubov's transformation:
 $ A^+=\alpha
 B^+\!+\!\beta^*B^-$,

6) postulate of Bogoliubov's vacuum : $B^-|0>_{U}=0$, and

7) cosmological creation of the
 {\it universes} from the Bogoliubov vacuum.

 Let us carry out this programme  in detail.

\subsection{Hamiltonian reduction}

 In the cosmological model (\ref{ham}), there are two independent
  equations: the one  of the lapse function
 $\delta S_{\rm cosmic}/\delta N_0=0$:
 \be\label{1conn}
 P^2_\vh=E_\vh^2,
 \ee
 treated as the  energy
 constraint, and the equation of momentum
 $\delta S_{\rm cosmic}/\delta P_\vh=0$
 \be\label{3conn}
 {P}_\vh= 2V_0\vh',
 \ee
  where $\vh'=\frac{d\vh}{d\eta}$.
 The constraint (\ref{1conn}) has two solutions
 \be\label{2conn}
 {P^{\pm}_\vh}=\pm E_\vh=\pm 2V_0\sqrt{\rho(\vh)},
 \ee
 where $E_\vh$, given by Eq. (\ref{ev1}), is identified with
 the ``one-universe energy''. The substitution of these solutions
 into action (\ref{ham}) and interval  (\ref{conf}) gives us
 their values
\be\label{ham-2}
 {S_{\rm cosmic}}_{|P_\vh=P^{\pm}_\vh}=S^{\pm}_{\rm cosmic}=\mp
2V_0\int\limits_{\vh}^{\vh_0}
 d\widetilde{\vh}\sqrt{\rho_0(\widetilde{\vh})},
 \ee
 and
\be\label{4conn}
 \eta(\vh|\vh_0)=2V_0
 \int\limits_{\vh}^{\vh_0}\frac{d\widetilde{\vh}}
 {{P^{\pm}_\vh}}=\pm
 \int\limits_{\vh}^{\vh_0}\frac{d\widetilde{\vh}}
 {\sqrt{\rho_0(\widetilde{\vh})}}\equiv\pm (\eta_0-r),
 \ee
 respectively. We called these values
 the {\it Hamiltonian reduction} of the geometric system
 (\ref{ham}), (\ref{conf}).
 Eq. (\ref{4conn}) is treated,
  in the observational cosmology \cite{039,Danilo},
  as the conformal version of the
  Hubble law. This law
   describes the relation between the redshift $z+1=\vh_0/\vh(\eta)$
  of spectral lines of photons (emitted by atoms on a cosmic object
  at the conformal time $\eta(\vh|\vh_0)=\eta$) and
  the coordinate distance $r=\eta_0-\eta$ of this object, where
  $\eta_0$ is the present-day moment. Thus, we see that
  WDW cosmology coincides with the Friedmann one
  $\vh{'}{}^2=\rho_0(\vh)$. Our task is to consider the status of this
 Friedmann cosmology in QFT of universes obtained by
 the first and the secondary quantization of the constraint
 (\ref{1conn}).

 \subsection{Quantum Field Theory of universes}

 After the primary  quantization of the cosmological scale factor $\vh$:
 $i[P_\vh,\vh]=1$ the energy constraint (\ref{1conn}) transforms
 to the WDW equation
\be\label{wdw}
 \partial^2_\vh\Psi+E_\vh^2\Psi=0.
 \ee
 This equation can be obtained in the corresponding classical
 WDW field theory for universes of the type of the Klein -- Gordon
 one:
 \be\label{uf}
 S_{\rm U}=\int d\vh \frac{1}{2}
 \left[(\partial_\vh\Psi)^2-E_\vh^2\Psi^2\right]\equiv \int d\vh
  L_{\rm U}.
 \ee
 Introducing the canonical momentum
$P_\Psi=\partial L_{\rm U}/\partial(\partial_\vh\Psi)$, one can
obtain the Hamiltonian form of this theory
 \be\label{ufh}
 S_{\rm U}=\int d\vh \left\{P_\Psi\partial_\vh\Psi-H_{\rm
 U}\right\},
 \ee
 where
\be\label{ufh1}
 H_{\rm U}=\frac{1}{2}\left[P_\Psi^2+E_\vh^2\Psi^2\right].
 \ee
 is the Hamiltonian. The concept of the one-universe ``energy'' $E_\vh$
 gives us the opportunity to present this Hamiltonian $H_{\rm U}$
 in the standard forms of the product of this ``energy''  $E_\vh$ and
 the ``number'' of universes
 \be\label{AA}
 N_U=A^+A^-,
 \ee
 \be\label{ufh1z12}
 H_{\rm U}=E_\vh\frac{1}{2}\left[A^+A^-+A^-A^+\right]=E_\vh[N_U+\frac{1}{2}]
 \ee
 by  means of the transition to the holomorphic variables
 \be\label{g}
 \Psi=\frac{1}{\sqrt{2E_\vh}}\{A^{+}+A^{-}\},~~~~~~~~
 P_\Psi=i\sqrt{\frac{E_\vh}{2}}\{A^{+}-A^{-}\}.
 \ee
 The  dependence of $E_\vh$  on $\vh$ leads to the additional term
 in the action expressed in terms the holomorphic variables
 \be\label{und3}
   P_\Psi \partial_\vh \Psi=
  \left[\frac{i}{2}(A^+\partial_\vh A^--A^+
 \partial_\vh A^-)-
 \frac{i}{2}(A^+A^+- A^-A^-)\triangle (\vh)\right],
\ee
 where
 \be\label{tri}
 \triangle(\vh)=\frac{\partial_\vh E_\vh}{2E_\vh}.
 \ee
  The last term in (\ref{und3}) is responsible for
  the cosmological creation of  ``universes'' from ``vacuum''.

 \subsection{Bogoliubov transformation and creation of universes}

 In order to define stationary physical states, including
 a ``vacuum'', and a set of integrals of motion, one usually uses
 the Bogoliubov transformations \cite{B,ps1} of the holomorphic
 variables of universes
 $(A^+,A^-)$:
 \be \label{u17} A^+=\alpha
 B^+\!+\!\beta^*B^-,~~~~~~~\;\;A^-=\alpha^*
 B^-\!+\!\beta A^+~~~~~~~~~~~(|\alpha|^2-|\beta|^2=1),
  \ee
  so that the classical equations of the field theory  in terms of universes
 \be \label{1un} (i\partial_\vh+E_\vh)A^+=iA^-\triangle(\vh),~~~~~~~~
(i\partial_\vh-E_\vh)A^-=iA^+\triangle(\vh),
  \ee
  take a diagonal form in terms of  {\it quasiuniverses}
  $B^+,B^-$:
 \be \label{2un} (i\partial_\vh+E_B(\vh))B^+=0,~~~~~~~~
 (i\partial_\vh-E_B(\vh))B^-=0.
  \ee
  The diagonal form is possible, if the  Bogoliubov coefficients $\alpha,\beta$ in
  Eqs. (\ref{u17})
  satisfy to equations
 \be \label{3un} (i\partial_\vh+E_\vh)\alpha=i\beta\triangle(\vh),~~~~~~~~
(i\partial_\vh-E_\vh)\beta^*=i\alpha^*\triangle(\vh).
  \ee
 For the parametrization
 \be \label{4un} \alpha=e^{i\theta(\vh)}\cosh r(\vh),~~~~~~~~
 \beta^*=e^{i\theta(\vh)}\sinh{r}(\vh),
  \ee
  where $r(\vh),\theta(\vh)$ are the parameters of ``squeezing''
  and ``rotation'', respectively, Eqs. (\ref{3un}) become
 \be \label{5un}
 (i\partial_\vh\theta-E_\vh)\sinh 2r(\vh)=-\triangle(\vh)\cosh 2r(\vh)\sin
 2\theta(\vh),~~~~~~~~\partial_\vh r(\vh) =\triangle(\vh)\cos
 2\theta(\vh),
 \ee
  while ``energy'' of  {\it quasiuniverses} in Eqs. (\ref{2un})
  is defined by expression
   \be \label{6un}
  E_B(\vh)=\frac{E_\vh-\partial_\vh\theta(\vh)}{\cosh 2r(\vh)}.
  \ee
 Due to Eqs. (\ref{2un}) the ``number'' of {\it quasiuniverses}
  ${\cal N}_B=(B^+B^-)$ is conserved
  \be \frac{d{\cal N}_B}{d\vh}\equiv
  \frac{d(B^+B^-)}{d\vh}=0.
  \ee
  Therefore, we can introduce the ``vacuum''
 as a state without {\it quasiuniverses}:
 \be \label{sv}
 B^-|0>_{\rm U}=0.
 \ee
 A number of created {\it universes} from this Bogoliubov vacuum
 is equal to the expectation value of the operator
 of the  {\it number of universes } (\ref{AA}) over the Bogoliubov
 vacuum
 \be\label{usv1}
 N_{\rm U}(\vh)={}_{\rm U}<A^+A^-
  >_{\rm U}\equiv |\beta|^2=\sinh^2r(\vh),
 \ee
 where $\beta$ is the coefficient in the Bogoliubov transformation
 (\ref{u17}), and $N_{\rm U}(\vh)$ is called the ``distribution
 function''. Introducing the Bogoliubov ``condensate''
\be\label{usv2}
 R_{\rm U}(\vh)=i(\alpha\beta^*-\alpha^*\beta)
 \equiv {}_{\rm U}<P_\Psi\Psi >_{\rm U}
 =\frac{i}{2}~\,{}_{\rm U}\!<[A^+A^+-A^-A^-]>_{\rm U},
 \ee
 one can rewrite the Bogoliubov equations of the diagonalization (\ref{3un})
\be\label{usv3}
 \left\{\begin{aligned}
 \frac{dN_{\rm U}}{d\vh}&=\triangle(\vh)
 \sqrt{4N_{\rm U}(N_{\rm U}+1)-R_{\rm U}^2},  \\
 \frac{dR_{\rm U}}{d\vh}&=-{2E_\vh}
 \sqrt{4N_{\rm U}(N_{\rm U}+1)-R_{\rm U}^2}.
 \end{aligned}\right.
 \ee
  It is natural to propose that at the moment of creation of
  the universe $\vh(\eta=0)=\vh_I$ both these functions are equal to zero
   $N_{\rm U}(\vh=\vh_I)=R_{\rm U}(\vh=\vh_I)=0$.
  This moment of the conformal time (\ref{4conn})
  $\eta=0$
   is distinguished by the vacuum postulate (\ref{sv}) as the beginning
   of a universe.

\subsection{Quantum anomaly of conformal time}

 As it was shown in the case of a particle in QFT \cite{Logunov},
 the postulate of a vacuum as a state with minimal ``energy''
 restricts the motion of a ``universe'' in
 the space of events, so that a ``universe'' with $P_{\vh+}$ moves
 forward and with $P_{\vh-}$ backward.
 \be \label{1b12sr}
 P_{\vh+}~~~~ \to ~~~~\vh_{I}\leq {\vh_{0}};~~~~~~~~~~~~~~~~~~
 ~~~P_{\vh-}~~~~ \to ~~~~\vh_{I}\geq {\vh_{0}}.
  \ee
  If we substitute this restriction into the interval (\ref{4conn})
 \be  \label{u7+}
 \eta_{(P_{vh+})}=\int\limits_{\vh_I}^{\vh_0}\frac{d\vh}{\sqrt{{\rho_0(\vh)}}}
 ;~~~~~~~~~~~~~~~~~~\vh_{I}\leq {\vh_{0}},
 \ee
 \be  \label{u7-}
 \eta_{(P_{\vh-})}=\int\limits_{\vh_0}^{\vh_I}\frac{d\vh}{\sqrt{\rho_0(\vh)}}
 ;~~~~~~~~~~~~~~~~~~\vh_{I}\geq {\vh_{0}},
 \ee
 one can see that the geometric interval in both cases is
 positive. In other words, the stability of quantum theory
 as the vacuum postulate leads to the absolute point of
 reference of this interval $\eta=0$  and its positive arrow.
   In QFT the initial datum $\vh_I$ is considered  as a point of
 creation or annihilation of universe.
 One can propose that the singular point $\vh=0$ belongs to
 antiuniverse. In this  case, a universe with a positive
 energy goes out of the singular point   $\vh =0$.

 In the model of rigid state $\rho=p$, where  $E_\vh=Q/\vh$
 Eqs. (\ref{usv3}) have an exact solution
\be\label{11cu}
 N_{\rm U}=\frac{1}{4Q^2-1}
 \sin^2\left[\sqrt{Q^2-\frac{1}{4}}~~\ln\frac{\vh}{\vh_I}\right]\not
 =0,
\ee
 where
 \be\label{cc}
 \vh=\vh_I\sqrt{1+2H_I\eta}
 \ee
  and
$\vh_I,H_I=\vh'_I/\vh_I=Q/(2V_0\vh_I^2)$ are the initial data.

 We see that there are results of the type of the arrow of
 time and absence of the cosmological singularity (\ref{u7+}),
 which can be understood only on the level of
  quantum theory, where symmetry $\eta~\to~-\eta$ is broken \cite{riv,ilieva}.


\section{General Relativity}

     General Relativity (GR)  \cite{einsh,H}
    is given by two fundamental quantities:
    the  {\it``dynamic''} action (\ref{h1})  and
 {\it``geometric interval''}  (\ref{h2})
\be \label{ds}
 g_{\mu\nu}dx^\mu dx^\nu\equiv\omega_{(\alpha)}\omega_{(\alpha)}=
 \omega_{(0)}\omega_{(0)}-
 \omega_{(1)}\omega_{(1)}-\omega_{(2)}\omega_{(2)}-\omega_{(3)}\omega_{(3)},
 \ee
 where are $\omega_{(\alpha)}$ linear differential forms introduced
 by Fock  \cite{fock29} as components of an orthogonal
 simplex of reference.

 Hilbert's
 {\it``Foundation of Physics''}
 in terms of Fock's  simplex 
  contains  two principles of
 relativity: the {\it``geometric''} --- general coordinate transformations
\be \label{1zel}
 x^{\mu} \rightarrow  \tilde x^{\mu}=\tilde
 x^{\mu}(x^0,x^{1},x^{2},x^{3}),~~~~~~~~~
 \omega_{(\alpha)}(x^{\mu})~\to ~\omega_{(\alpha)}(\tilde x^{\mu})=
 \omega_{(\alpha)}(x^{\mu})
 \ee
 and the {\it``dynamic''} principle formulated as the Lorentz
 transformations of an orthogonal  simplex of reference
 \be \label{2zel}
{\omega}_{(\alpha)}~\to ~
\overline{\omega}_{(\alpha)}=L_{(\alpha)(\beta)}{\omega}_{(\beta)}.
\ee
  The latter are considered as transformations of a frame of reference.

  Fock's  separation of  the frame transformations  (\ref{2zel})
 from the gauge ones (\ref{1zel})
  \cite{fock29} allows us to consider GR and SR on equal footing.
A direct pathway from Hilbert's
 geometric formulation of GR to Quantum Gravity
 passed by
 contemporary QFT
through Dirac's Hamiltonian reduction   and Bogoliubov
 transformations \cite{B,ps1}:

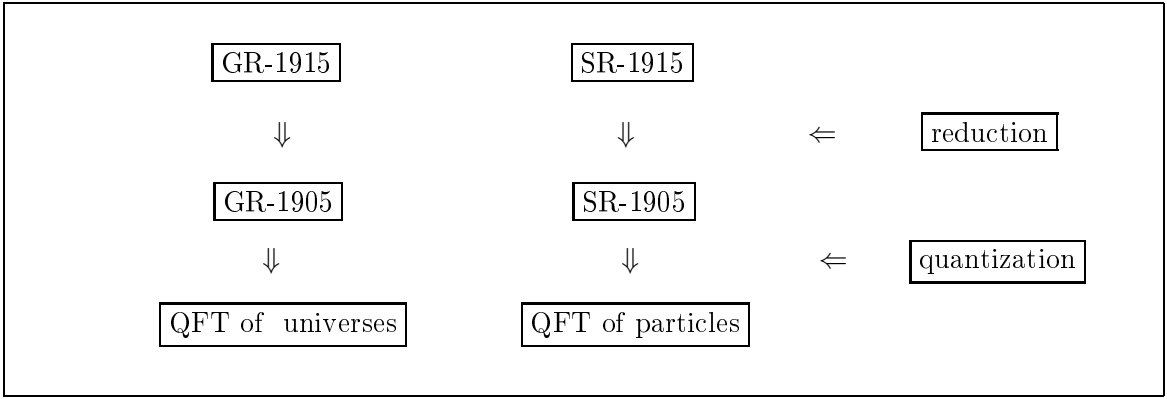
\vspace{4mm}

 \begin{tabular}{|p{15cm}|}
 \hline

 \vspace{2mm}
\label{gd} ~~~~~~~~~~~~~~~~~~~~\fbox{\rm
GR-1915}~~~~~~~~~~~~~~~~~~~~~~~~\fbox{\rm SR-1915}
 ~~~~~~~~~~~~
 $$
 ~~~~~~~~~~~~~~~~~~~~~~~ ~\Downarrow~~~~~~~~~~~~~~~~~~~~~~~~ ~~~~~~~~~~ \Downarrow
 ~~~~~~~~~~~~~~~~~~\Leftarrow~~~~~~~~\fbox{\rm reduction}~~~~~~~
 $$
 $$
\fbox{\rm GR-1905}~~~~~~~~~~~~~~~~~~~~~~~~\fbox{\rm
SR-1905}~~~~~~~~~~~~~~~~~~~~~~~~~~~
 $$
$$
 ~~~~~~~~~~~~~~~~~~~~~~~~~~~\Downarrow ~~~~~~~~~~~ ~~~~~~~~~~~~~~~ ~~~~~~~~~~~~\hfill \Downarrow
 ~~~~~~~~~~~~~~~~~~~~\Leftarrow~~~~~~\fbox{\rm
 quantization}~~~~~~~
$$
$$
~~~~~~~~~~~\fbox{\rm QFT~of~ universes}~~~~~~~~~
 ~~~\fbox{\rm QFT~of~particles}~~~~~~~~~~~~~~~~~~~~~~~~~~~~~~
 ~~~~~~~~
$$\\
\hline
\end{tabular}

\vspace{4mm}

 was reconstructed in \cite{pvng8b}.

\subsection{\label{adm1}The Dirac -- ADM frame}

  The Hamiltonian approach to GR is formulated
  in the frame of reference  given by  Fock's simplex of reference
  in terms of the Dirac variables \cite{dir}
\be \label{adm}
 \omega_{(0)}=\psi^6N_{\rm d}dx^0,~~~~~~~~~~~
 \omega_{(b)}=\psi^2 {\bf e}_{(b)i}(dx^i+N^i dx^0);
 \ee
 here triads ${\bf e_{(a)i}}$ form the spatial metrics with $\det |{\bf
 e}|=1$.
\subsection{\label{ze}Zel'manov's class of functions of
  diffeomorphisms}

  Zel'manov established in \cite{vlad} that a frame of reference determined by
   forms (\ref{adm}) is invariant with respect to
  transformations
 \be \label{zel}
 x^0 \rightarrow \tilde x^0=\tilde x^0(x^0);~~~~~
 x_{i} \rightarrow  \tilde x_{i}=\tilde x_{i}(x^0,x_{1},x_{2},x_{3})~,
 \ee
 \be \label{kine}
 \tilde N_d = N_d \frac{dx^0}{d\tilde x^0};~~~~\tilde N^k=N^i
 \frac{\partial \tilde x^k }{\partial x_i}\frac{dx^0}{d\tilde x^0} -
 \frac{\partial \tilde x^k }{\partial x_i}
 \frac{\partial x^i}{\partial \tilde x^0}~.
 \ee
 This group of transformations conserves
  a family  of hypersurfaces  $x^0=\rm{const.}$,
  and it calls the {\it``kinemetric''} subgroup of the group of
  general coordinate  transformations.
    The {\it``kinemetric''} subgroup contains
 reparametrizations of the coordinate evolution parameter.

 The reparametrization of the coordinate evolution parameter $(x^0)$
   means that this specific frame of reference
  (\ref{adm}) should be redefined by pointing out two {\it Dirac observables}:
{\it time-like variable} in the {\it field space of events} and
the {``time''} as a {\it geometric
 interval} \cite{ps1}.

\subsection{\label{s-3}Separation of  Scale Factor}

The cosmological evolution is
 the irrefutable observational
 argument in favor of
 existence of such a homogeneous variable  considered in GR
  as the cosmological scale factor.
  The  cosmological
 scale factor  $a(x_0)$ introduced
 by the scale transformation:
 $g_{\mu\nu}=a^2(x_0)\widetilde{g}_{\mu\nu}$, where $\widetilde{g}_{\mu\nu}$
 is defined by (\ref{adm}), where
 $\widetilde{N}_d=a^{2}{N}_d$ and
 $\widetilde{\psi}^2=a^{-1}\!\psi^2.$
 In  order to keep the number of variables of GR, the scale factor
can be defined using the spatial averaging $ \log
\sqrt{a}\equiv\langle{\log {\psi}}\rangle
 \equiv V_0^{-1}\int d^3x \log {\psi}$, so that the rest
 scalar component $\widetilde{\psi}$
  satisfies  the identity \cite{pvng8b}
\be\label{non1}
 \langle\log\widetilde{\psi}\rangle
 \equiv V_0^{-1}\int d^3x \log\widetilde{\psi}
 =
 V_0^{-1}\int d^3x \left[\log{\psi}
 -\left\langle{ \log{\psi}}\right\rangle\right]\equiv 0,
 \ee
 where $V_0=\int d^3x$ is finite volume.
 The similar scale transformation of a curvature
 $
 \sqrt{-g}R(g)=a^2\sqrt{-\widetilde{g}}R(\widetilde{g})-6a
 \partial_0\left[{\partial_0a}\sqrt{-\widetilde{g}}~\widetilde{g}^{00}\right]$
  converts action (\ref{h1}) into
 \be\label{1gr}
 S=\widetilde{S}-
 \int\limits_{V_0} dx^0 (\partial_0\vh)^2\int {d^3x}{\widetilde{N}_d}^{-1},
 \ee
 where $\widetilde{S}$
  is the action (\ref{h1})  in
 terms of metrics $\widetilde{g}$ and
 the running scale  of all masses
 $\vh(x^0)=\vh_0a(x^0)$   and $(\widetilde{N}_d)^{-1}=
 \sqrt{-\widetilde{g}}~\widetilde{g}^{00}$.
 One can construct the Hamiltonian function using
 the definition of a set of  canonical
 momenta:
\bea \label{pph}
 P_\vh&=&\frac{\partial L}{\partial (\partial_0\vh)}
 = -2V_0\partial_0\vh
 \left\langle(\widetilde{N}_d)^{-1}\right\rangle=
 -2V_0\frac{d\varphi}{d\zeta}\equiv-
2V_0 \vh',
 \\
 \label{gauge}
 p_{\psi}&=&\frac{\partial {\cal L}}{\partial (\partial_0\log\widetilde{\psi})}\equiv
 -\frac{4\vh^2}{3}\cdot\frac{\partial_l(\widetilde{\psi}^{6}N^l)-
 \partial_0(\widetilde{\psi}^{6})}{\widetilde{\psi}^{6}\widetilde{N_d}},
 \eea
where $d\zeta=\langle(\widetilde{N}_d)^{-1}\rangle^{-1}dx^0$ is a
time-interval invariant with respect to
  time-coordinate transformations
  $x^0 \to \widetilde{x}^0=\widetilde{x}^0(x^0)$.
The  Hamiltonian form of the action  in terms of momenta $P_\vh$
and $P_{ F}=[{p_{\psi}}, p^i_{{(b)}},p_f]$ including (\ref{pph}),
(\ref{gauge}) takes the form
 \be\label{hf}
S=\int dx^0\left[\int d^3x \left(\sum_F P_F\!\partial_0
F\!+\!C\!-\!\widetilde{N}_d\widetilde{T}^0_0\right)\!-\!P_{\varphi}\partial_0\varphi+
\frac{P_{\varphi}^2}{4\int dx^3 ({\widetilde{N}_d})^{-1}}\right],
\ee where
 ${\cal C}=N^i {T}^0_{i} +C_0p_{\psi}+ C_{(b)}\partial_k{\bf
e}^k_{(b)}$
  is the sum of constraints
  with the Lagrangian multipliers $N^i,C_0,~C_{(b)}$ and the energy--momentum tensor
  components $T^0_i$; these constraints include
   the transversality  $\partial_i {\bf e}^{i}_{(b)}\simeq 0$ and the Dirac
 minimal  surface \cite{dir}:
 \be\label{hg}
{p_{\psi}}\simeq 0 ~~~~\Rightarrow ~~~~
\partial_j(\widetilde{\psi}^6{\cal N}^j)=(\widetilde{\psi}^6)'
~~~~~ ({\cal N}\,^j=N^j\langle \widetilde{N}_d^{-1}\rangle).
 \ee
The explicit dependence of $\widetilde{T}_0^0$ on
$\overline{\psi}$
  can be given in terms of the scale-invariant  Lichnerowicz
  variables \cite{Y} in the space-time defined by the Fock simplex
  $\omega^{(L)}_{(\mu)}=\psi^{-2}\omega_{(\mu)}$:
 \be\label{t00}
 \widetilde{T}^0_0= \widetilde{\psi}^{7}\hat \triangle \widetilde{\psi}+
  \sum_I \widetilde{\psi}^Ia^{I/2-2}\tau_I, \ee
   where $\hat \triangle
 F\equiv({4\varphi^2}/{3})\partial_{(b)}\partial_{(b)}F$ is
 the Laplace operator and
  $\tau_I$ is partial energy density
  marked by the index $I$ running a set of values
   $I=0$ (stiff), 4 (radiation), 6 (mass), 8 (curvature), 12
   ($\Lambda$-term)
in accordance with a type of matter field contributions, and $a$
is the scale factor.

 \subsection{Hamiltonian Reduction}

 The
 energy constraint ${\delta S[\vh_0]}/{\delta  \widetilde{N_d}}=0$ takes the
 algebraic form
 \be\label{nph}
 -
 \frac{\delta \widetilde{S}[\vh]}{\delta  \widetilde{N}_d}
 \equiv \widetilde{T}^0_0=\frac{(\partial_0\varphi)^2}{\widetilde{N}_d^2}
 =\frac{P_\vh^2}{4V_0^2[{\langle(\widetilde{N}_d)^{-1} \rangle
 \widetilde{N}_d}]^2},
 \ee
 where $T^0_0$ is the local energy density by definition.
The spatial averaging
  of this equation multiplied by $\widetilde{N}_d$ looks like the energy constraint
 \be \label{ec}
 P^2_{\varphi}=E^2_{\varphi},
 \ee
 where the  Hamiltonian functional $ E_\vh=2\int
 d^3x(\widetilde{T}^0_0)^{1/2}= 2V_0{\langle
 (\widetilde{T}^0_0)^{1/2}\rangle}
 $
 can be treated as the ``universe energy'' by analogy with the ``particle energy'' in
 special relativity (SR).
 Eqs. (\ref{pph}) and (\ref{ec}) have the exact solution
 \be\label{13c}
 \zeta(\varphi_0|\varphi)
 \equiv\int dx^0 {\left\langle
 (\widetilde{N}_d)^{-1}\right\rangle}^{-1}
 =\pm
 \int_{\vh}^{\vh_0}
 {d\widetilde{\vh}}{{\langle
 (\widetilde{T}^0_0(\widetilde{\vh}))^{1/2}\rangle}}^{-1}
 \ee
 well known as the Hubble-type evolution.
 The local part of Eq. (\ref{nph}) determines
 a   part of
 the Dirac lapse function invariant with respect to
 diffeomorphisms  of the
Hamiltonian formulation $x^0 \to
\widetilde{x}^0=\widetilde{x}^0(x^0)$ \cite{vlad}:
 \be\label{13ec}
 N_{\rm inv}={\langle(\widetilde{N}_d)^{-1} \rangle
 \widetilde{N}_d}={{\left\langle\sqrt{{\widetilde{T}^0_0}}\right\rangle}}
 \left(\sqrt{{\widetilde{T}^0_0}}\right)^{-1}.
 \ee

 One can find
 evolution of all field variables $F(\vh,x^i)$  with respect to
 $\vh$ by the variation of the ``reduced'' action obtained as
   values of the Hamiltonian form of the initial action  (\ref{hf}) onto
 the energy constraint  (\ref{ec}) \cite{pvng8b}:
 \be\label{2ha2} S|_{P_\vh=\pm E_\vh} =
 \int\limits_{\vh}^{\vh_0}d\widetilde{\vh} \left\{\int d^3x
 \left[\sum\limits_{  F}P_{  F}\partial_\vh F
 +\bar{\cal C}\mp2\sqrt{\widetilde{T}_0^0(\widetilde{\vh})}\right]\right\},
\ee
 where $\bar{\cal C}={\cal
 C}/\partial_0\widetilde{\vh}$.  The reduced Hamiltonian
 $\sqrt{\widetilde{T}_0^0}$ is
 Hermitian,
 if the  minimal surface
 constraint
 (\ref{hg}) removes a negative
 contribution of $p_{\psi}$ from the energy density.
 The reduced action (\ref{2ha2}) shows us
 that the initial data at the beginning $\vh=\vh_I$ are independent of
  the present-day ones at  $\vh=\vh_0$,
  therefore
  the proposal about an existence of the  Planck epoch $\vh=\vh_0$
   at the beginning \cite{bard} looks
  very doubtful. Let us consider consequences of
  the classical reduced theory (\ref{2ha2}) and quantization of
  the energy constraint (\ref{ec}) without the ``Planck epoch''
  at the beginning.

\subsection{\label{s-4}The  low-energy decomposition
  of ``reduced''  action}

We have seen above that in the ``reduced''  action (\ref{2ha2})
momenta
 ${P_\vh}_{\pm}=\pm E_\vh$
   become  the generators of evolution of all variables with respect to the
  evolution parameter $\vh$  forward and backward,
 respectively.

 Let us consider the positive branch and assume that the local density
 $T_0^0=\rho_{(0)}(\vh)+T_{\rm f}$
 contains a tremendous  cosmological background
$\rho_{(0)}(\vh)$.
 The  low-energy decomposition
  of ``reduced''  action (\ref{2ha2})  $2 d\vh \sqrt{\widetilde{T}_0^0}= 2d\vh
\sqrt{\rho_{(0)}+T_{\rm f}}
 =
 d\vh
 \left[2\sqrt{\rho_{(0)}}+
 T_{\rm f}/{\sqrt{\rho_{(0)}}}\right]+...$
 over
 field density $T_{\rm f}$ gives the sum
 $S|_{P_\vh=+E_\vh}=S^{(+)}_{\rm cosmic}+S^{(+)}_{\rm
 field}+\ldots$, where the first  term of this sum
 $S^{(+)}_{\rm cosmic}= +
 2V_0\int\limits_{\vh_I}^{\vh_0}\!
 d\vh\!\sqrt{\rho_{(0)}(\vh)}$ is  the reduced  cosmological
 action,
 whereas the second one is
  the standard field action of GR and SM
 \be\label{12h5} S^{(+)}_{\rm field}=
 \int\limits_{\zeta_I}^{\zeta_0} d\zeta\int d^3x
 \left[\sum\limits_{ F}P_{ F}\partial_\eta F
 +\bar{{\cal C}}-T_{\rm f} \right]
 \ee
 in the  space determined by the interval
 \be\label{d2}
 ds^2=d\zeta^2-\sum_a[e_{(a)i}(dx^i+{\cal N}^id\zeta)]^2;
 ~~\partial_ie^i_{(a)}=0,~~\partial_i{\cal N}^i=0
 \ee
 with  conformal time
 $d\eta=d\zeta=d\vh/\rho_{(0)}^{1/2}$ as the diffeo-invariant
 and scale-invariant quantity, coordinate distance
 $r=|x|$,
 and running masses
 $m(\zeta)=a(\zeta)m_0$.
 This expansion shows us that the Hamiltonian approach
 identifies the ``conformal quantities''
  with the observable ones including the conformal time $d\eta$,
  instead of $dt=a(\eta)d\eta$, the coordinate
 distance $r$, instead of Friedmann one $R=a(\eta)r$, and the conformal
 temperature $T_c=Ta(\eta)$, instead of the standard one $T$ \cite{039}.
 In this case
 the
  red shift of the spectral lines of atoms on cosmic objects
 $$
\frac{E_{\rm emission}}{E_0}=\frac{m_{\rm atom}(\eta_0-r)}{m_{\rm
atom}(\eta_0)}\equiv\frac{\vh(\eta_0-r)}{\vh_0}=a(\eta_0-r)
=\frac{1}{1+z}
$$
is explained by the running masses $m=a(\eta)m_0$ in action
(\ref{12h5}).

The conformal observable distance  $r$ loses the factor $a$, in
comparison with the nonconformal one $R=ar$. Therefore,
  the reduced interval (\ref{d2})
  describing the redshift --
  coordinate-distance relation \cite{039} corresponds to a different
  equation
  of state than in the case of the standard $\Lambda$- Dark Matter cosmology.
the Supernova data  are consistent with the dominance of the stiff
state, $\Omega_{\rm Stiff}\simeq 0.85 \pm 0.15$, $\Omega_{\rm
Matter}=0.15 \pm 0.10$ \cite{039,Danilo}, where we have the square
root dependence of the scale factor on conformal time
$a(\eta)=\sqrt{1+2H_0(\eta-\eta_0)}$. Just this time dependence of
the scale factor on
 the measurable time (here -- conformal one) is used for description of
 the primordial nucleosynthesis \cite{Danilo,three}.

 This stiff state is formed by a free scalar field
 when $E_\vh=2V_0\sqrt{\rho_0}=Q/\vh$. Just in this case there is an exact
solution of  Bogoliubov's equations (\ref{11cu})

\subsection{\label{s-5}The Quantum Universe}
  \label{ch-2}

  In the low-energy approximation GR is split onto the cosmological
  model considered in Section 2, and quantum field theory
  (\ref{12h5}) far from large masses
  \cite{pvng8b}. The wave function is a sum
   of products of the wave function of the
  homogeneous Universe and the wave function of the matter
  including gravitons settled in the space-time  described by (\ref{d2}) with
  $\widetilde{\psi}=1,N_{\rm int}=1$. The wave function of
  the matter fields with running masses is obtained by the
  Bogoliubov transformations (see in details \cite{ps1,114:a}).

\section{Conclusion}
 We gave here a set of theoretical and
 observational arguments in  favor of that
 General Relativity   has a consistent interpretation
 in the form of quantum theory of the type of the  microscopic
 theory of superfluidity \cite{Lon,Lan,B} with the Bogoliubov
 transformations used for construction of integrals of motion
 and stable physical states.

 Our hopes for an opportunity to construct a
 realistic quantum theory for GR
 were based, on the one hand, on the existence of
 the Hilbert-type geometric formulation \cite{H}  of Special
 Relativity  (SR) with the energy constraint
 considered  as the simplest model of GR
 and, on the other hand, on  the contemporary quantum field theory
 (QFT)  based on the primary and secondary
 quantization of this energy constraint
 \cite{Logunov}.

 We  reconstructed the direct pathway
 from geometry of GR \cite{H}
 to the  causal operator quantization of universes
 and to their quantum creation from vacuum considered as
  a state with the minimal ``energy''. This formulation included
 the Wheeler -- DeWitt definition of ``field space of events''
 \cite{WDW}, where diffeomorphisms were split
  from transformations of the
 frames of references using the Fock simplex of reference \cite{fock29};
 the choice
 of the Dirac specific frame of reference \cite{dir};
 resolving the energy constraint in the class of functions of the
 gauge transformations established by Zel'manov \cite{vlad};
  the calculation of values of the geometric action and interval
 onto the resolutions of the energy constraint \cite{pvng8b}
 in order to get the dynamic
 ``reduced'' action  in terms of difeo-invariant
 variables and to define the notions ``energy``,
 ``time``, ``particle'' and ``universe'',
  ``number'' of ``particles'' and ``universes'' by
    the low-energy expansion of this ``reduced'' action.

The  pathway of SR towards QFT \cite{Logunov} could be repeated
for GR, as the dimension of the diffeomorphism group of Dirac --
ADM Hamiltonian approach to GR coincided with the dimension of
constraints removing a part of canonical momenta in accordance
with the second N\"other theorem.
 This coincidence  takes place,
 if the cosmological evolution
 is considered as a collective dynamics \cite{pvng8b}. In this case,
 the preservation of
 the number of variables in GR leads to the Landau-type
 friction free  cosmic  motion \cite{Lan} with the London-type
 unique wave function \cite{Lon} and
 Bogoliubov-type transformations \cite{B,ps1}. The latter is used
   in order to obtain
 a set of integrals of motions and to calculate the distribution
 functions of cosmological creation of both ``universes'' and ``particles''
 from ``vacuum''. All these ``superfluid attributes''
  were accompanied by a set of physical consequences that could be
  understood as only pure quantum effects.
 In particular, this quantum dynamics of GR gave us
 possible solutions of the topical astrophysical problems,
 including horizon, homogeneity, cosmological singularity, arrow of
 time, CMB fluctuations \cite{pvng8b}.


 The Hamiltonian approach revealed the  double counting
 of the cosmological scale factor in
   the standard Lifshitz
 perturbation theory. It  means that this standard
 perturbation theory
  does not coincide with the Einstein theory \cite{pvng8b}.
Avoiding this  double counting, in order  to return back to GR,
 we have obtained new Hamiltonian  equations. These  equations do not contain
 the time derivatives
 that are responsible for the ``primordial power spectrum'' in
 the inflationary model \cite{bard}.
 However, Dirac's Hamiltonian approach to GR gave  us
 another  possibility  to explain this ``spectrum'' and
  other  problems of cosmology
  by  the cosmological creation of the primordial W-, Z- bosons
  from vacuum when
 their Compton length coincides with the universe horizon \cite{114:a}.



\vspace{1cm}

{\bf Acknowledgement}

\medskip
The authors are grateful to   B.M. Barbashov,  A.V. Efremov,
 E.A. Kuraev, N.M. Plakida, V.B. Priezzhev,
  and
 A.F. Zakharov for interesting and critical
discussions.



\begin{thebibliography}{99}

\bibitem{einsh}
{Einstein, A.}: Die Gr\"undlange der allgemeinen
Relativit\"atstheorie. {\it Ann. d. Phys.}\ {\bf 49} (1916)
769--826.
\bibitem{H} Hilbert, D.: {Die Grundlangen der Physik},
 {\it Nachrichten von der K\"on. Ges. der Wissenschaften zu G\"ottingen,
 Math.-Phys. Kl.}, Heft\ {\bf 3} (1915) 395--407.
\bibitem{WDW}
Wheeler, J.~A.  ``Superspace and the nature of quantum
geometrodynamics'', in {\it Batelle Recontres: 1967 Lectures in
Mathematics and Physics,} edited by C. DeWitt and J.A. Wheeler,
New York, 1968, 242-307; DeWitt, B.~C.  ``Quantum Theory of
Gravity. I. The Canonical Theory'', {\it Phys. Rev.} {\bf 160}
(1967) 1113--1148.
\bibitem{Logunov}  Bogoliubov, N. N., Logunov, A. A., Oksak, A. I., Todorov, I. T.,
{``Obshie Prinzipi Kvantovoj Teorii Polja''}, 1st edn.,  Nauka,
Moscow, 1987.
\bibitem{B}
Bogoliubov, N.~N.:  On the Theory of Superfluidity, {\it J. Phys.}
(USSR)\ {\bf 11} (1947) 23--32.


\bibitem{ps1}
Pervushin, V. N.  and Smirichinski, V. I., ``Bogoliubov
quasipaticles in constrained systems''{\it J. Phys. A: Math.
Gen.}, {\bf 32} (1999) 6191--6201.
%
\bibitem{Lan}
Landau, L.~D.:~ Theory of Superfluid Helium II, {\it ZhETF} \ {\bf
11} (1941) 592--611 (in Russian); {\it J. Phys. USSR}\  {\bf 5}
(1941) 71--90; Theory of the Superfluidity of Helium II, {\it PR}\
{\bf 60} (1941) 356--358.
\bibitem{Lon} London, F.:
The $\lambda$-Phenomenon of Liquid Helium and the Bose-Einstein
Degeneracy. {\it Nature}\  {\bf 141} (1938) 643--644.

\bibitem{pp} Pawlowski, M. and Pervushin, V.~N.:
Reparametrization-Invariant Path Integral in GR and "Big Bang" of
Quantum Universe,
 {\it Int. J. Mod. Phys.}\ {\bf A 16} (2001) 1715--1742; [hep-th/0006116].
\bibitem{039}
 Behnke, D.,  Blaschke, D.~B.,  Pervushin, V.~N. and Proskurin,
 D.~V.:
  Description of supernova data in conformal
cosmology without cosmological constant,
 {\it Phys. Lett.}  \ {\bf B 530} (2002) 20--26; [gr-qc/0102039].

\bibitem{Danilo}
Behnke, D. (2004) Conformal Cosmology Approach to the Problem of
Dark Energy,
 PhD Thesis,
Rostock Report MPG-VT-UR 248/04.
\bibitem{riv}
Pervushin, V. N.: On the Vacuum in the Gauge Theories,  {\it Riv.
del Nuovo Cimento} \ {\bf 8} N 10 (1985) 1--59 .
\bibitem{ilieva}
 Ilieva, N. and Pervushin, V. N.:
 Gauge-Field Topology in
Two-Dimensions: $\theta$-Vacuum, Topological Phase, and Composite
Fields,  {\it Int. Journ. Mod. Phys.} \ {\bf 6} (1991) 4687--4697.





\bibitem{fock29}
Fock, V.~A.  ``Geometrisierung der Diracschen Theorie des
Electrons'', {\it Zs. Phys.} {\bf 57} (1929) 261--277.

\bibitem{dir}
Dirac, P.~A.~M.  ``Generalized Hamiltonian Dynamics'',
 {\it Proc. Roy. Soc. (London)} {\bf A 246} (1958) 326--332.


\bibitem{vlad}
Zelmanov, A.~L.  ``Kinemetric invariants and their relation to
chronometric ones in Einstein's gravitation theory'', {\it Dokl.
AN USSR} {\bf 209} (1973) 822--825.

\bibitem{pvng8b} Barbashov, B.M., Pervushin, V.N., Zakharov, A.F., and
Zinchuk, V.A.,
  ``Hamiltonian Cosmological Perturbation Theory'',
   {\it Phys. Lett.} ~in press (2006); {[hep-th/0501242]}.



\bibitem{Y}
York, J.~W.   ``Gravitational Degrees of Freedom and the
Initial-Value Problem'',
 {\it Phys. Rev. Lett.} {\bf 26} (1971) 1656--1658.


\bibitem{bard}
Mukhanov, V.~F., Feldman H.~A. and Brandenberger, R.~H. Theory of
cosmological perturbations, {\it Phys. Rep.} {\bf 215} (1992)
203--333.

\bibitem{three} {Weinberg, S.}
``First Three Minutes. A modern View of the Origin of the
universe'',
Basic Books, Inc., Publishers, New-York, 1977.


\bibitem{114:a}
Blaschke, D.~B. et~al.
``Cosmological Production of Vector
Bosons and Cosmic Microwave Background Radiation'', {\it Physics
of Atomic Nuclei} {\bf 67} (2004)
 1050--1062; [hep-ph/0504225].


\end{thebibliography}
\end{document}